\documentstyle [preprint,prd,aps] {revtex}
\begin{document}
\draft
\title {Exclusive photoproduction of $\Phi$ on Proton in the
quark-diquark model}

\author { C. Carimalo$^{(a)}$\footnote{e-mail : carimalo@in2p3.fr}, N.
Arteaga-Romero
$^{(a)}$, S. Ong$^{(b)}$\footnote{e-mail : ong@ipno.in2p3.fr}}

\address{$^{a)}$Coll\`ege de France, IN2P3-CNRS, Laboratoire de Physique
Corpusculaire,
11, place Marcelin Berthelot, F-75231 Paris Cedex 05, France\\
$^{b)}$ Institut de Physique Nucl\'eaire, IN2P3-CNRS, Universit\'e
Paris-Sud, 91406 Orsay Cedex, France}

\maketitle
\vskip 10 true cm
\begin{center}
{\bf {Abstract} }
\end{center}

We present predictions for exclusive photoproduction of 
$\Phi$-meson on proton at large transfer, where we use a quark-diquark 
structure model for the proton. Extrapolation from our results to lower 
transfers is comparable in magnitude with available data in that range. 
This may support the diquark model in its ability to provide, for that 
process, an appropriate link between diffractive physics at low transfer 
and the standard semi-perturbative approach of hard exclusive processes at 
very large transfer, where the proton recovers its three-quark structure.\\

\newpage

\section{  Introduction }

Up to now, exclusive photoproduction of a vector meson $V$ ($V = \rho, 
\omega, \Phi, J/\Psi$) from proton (reaction (1)) have been measured 
mainly at very low values of $t$, $t$ being the opposite of 
the squared momentum transfer at the proton vertex.

\begin{equation}
\gamma + p \rightarrow V + p
\end{equation}

In that region (say, $t \leq 1$ GeV$^2$), and in a wide energy range 
up to HERA energies, the observed characteristics of photoproduction 
of the lighter mesons are those of a soft diffractive process. 
As shown by Donnachie and Landshoff \cite{1}, this can be well described 
for the most part in a picture using both Vector Dominance Model 
(VDM) and Pomeron phenomenology : there, the incoming photon 
is assumed to convert into a vector meson which afterwards exchanges 
a soft Pomeron with the proton target. In this respect, one may consider 
reaction (1) in this small-transfer range as a good testing bench 
for Pomeron Physics. Indeed, that picture works nicely 
in case of photoproduction 
of the lighter mesons $\rho$, $\Phi$ and $\omega$ \cite{2}. However, 
it fails to reproduce the energy dependence of the cross section 
for $J/\Psi$ photoproduction \cite{3}. 

In a QCD-inspired picture, Pomeron exchange is commonly modelled as 
the effective exchange of two non-perturbative gluons. Donnachie and 
Landshoff improved substantially that picture \cite{4}. 
When applied to photoproduction of light mesons, their 
two-gluon exchange model leads to results very similar to those 
provided by the Pomeron-exchange model of the same authors \cite{5}. 
However, it also fails to describe the energy 
dependence of $J/\Psi$ photoproduction, 
as well as those of virtual photoproductions of $\rho$, $\Phi$ 
and $J/\Psi$, as observed at HERA \cite{6}.

Following the works of Ryskin and of Brodsky et al. \cite{7}, 
this may reveal that QCD perturbative 
effects enter the game, since a large momentum scale 
(either the mass of the heavy vector meson produced 
in case of $J/\Psi$, or the high $Q^2$ of the virtual photon 
in case of $\rho$) appears in the reaction. In their 
approach, Brodsky et al. wrote the amplitude of the process 
as the product of three \mbox{terms :} an amplitude 
describing the breaking of the virtual photon into 
a $q{\bar q}$ pair, the valence $q {\bar q}$ wave function of the 
vector meson and an amplitude describing a non-perturbative 
two-gluon structure of the proton. Actually, the latter amplitude plays 
a crucial role in this approach since the dependence on energy of the cross 
section directly reflects the small-$x$ behaviour of the gluon 
momentum distribution 
in the proton. In this way, one can account for the 
rapid rise with energy of the cross sections, observed at HERA.

In this paper, we consider (real) photoproduction of $\Phi$ 
on proton at larger $t$. Since the $\Phi$-meson is a pure 
$s {\bar s}$ state and the strangeness content of the nucleon 
wave function is probably small, that process is dominated, 
at lowest order in QCD, by the two-gluon exchange mechanism 
and thus provides a unique way to study the latter. This process has 
been already investigated at moderate $t$ by Laget and Mendez-Galain \cite{8}, 
using the non-perturbative picture 
of Donnachie and Landshoff. Their prediction for the 
differential cross section $d \sigma /dt$ at infinite photon energy 
exhibits a node at $t= 2.4 
$GeV$^2$, which, in turn, could serve as a test of the model. 
Unfortunately, the only existing data for $\Phi$-production 
with real photons or virtual photons correspond to very low values 
of $t$. 

Since higher values of $t$ provide larger momentum scales, 
it is tempting to apply in that range the semi-perturbative 
approach of hard exclusive processes developped long ago by Brodsky, 
Farrar, Lepage (BFL) \cite{9}, and by Chernyak and Zhitnitsky (CZ) \cite{10}. 
This approach has been described and discussed many times in the 
literature, and shown to provide a correct order of 
magnitude of numerous exclusive amplitudes. Indeed, Farrar et 
al. already applied it to various photoproduction processes \cite{11}.
 
We just remind here that in this formalism, the amplitude of a given 
exclusive process is obtained by a convolution formula of relativistic 
hadron wave functions with elementary hard scattering amplitudes 
involving the valence quarks and antiquarks of the hadrons 
taking part in the reaction. When using the baryon wave functions 
derived from QCD sum rules by (CZ), this 
method provides a correct order of magnitude 
for many exclusive amplitudes involving baryons. However, 
there exist important subasymptotic helicity-flip 
effects that do not fit the above picture where any spin 
effect is to be described by the so-called helicity conservation rule
\cite{12}. 
While this rule should be valid asymptotically, it appears to be 
inconsistent with most experimental data at intermediate energies. 
To cure this failure for processes involving baryons, a quark-diquark 
structure of baryons has been proposed \cite{13}. In that alternative 
picture, two of 
the three quarks of a baryon are clustering together in a diquark 
structure. In the subasymptotic region, diquarks are supposed to 
act as quasi-elementaries constituents having direct couplings 
with photons and gluons. Helicity-flips are then 
caused by vector diquarks. On the other hand, diquarks should 
asymptotically dissolve into quarks, restoring in this way the 
usual three-quark picture of baryons.
The diquark hypothesis provides natural explanations for many phenomena 
that are otherwise difficult to describe by standard models \cite{14}. 
In the following, 
we apply that picture in the framework of the (BFL) scheme, as 
a first semi-perturbative calculation of the process under study, 
from moderate to large values of $t$.

Another point of theoretical interest in the study of elastic 
$\Phi$-photoproduction is found in the structure of the 
amplitude of the underlying hard scattering process. 
Indeed, that amplitude exhibits singularities coming from on-shell 
quark lines. Farrar et al. \cite{15} have shown that in fact, any exclusive 
photoproduction process is, to leading twist, 
insensitive to long distance physics 
and do not require Sudakov resummation. The propagator singularities 
are integrable and their presence does not affect the validity of the hard 
scattering approach. The appearance of an imaginary part of the amplitude 
at leading order in $\alpha_s$ is thus considered as a non trivial 
prediction of perturbative QCD.

Studies of $J/\Psi$ or ${\eta}_c$ photoproduction are of course of the 
same interest as that of $\Phi$ photoproduction, since the charm content 
of the proton is probably negligible too, or even inexistent. 
However, on one hand, the high value of the c-quark mass is a source of 
computational complications, and, in the other hand, more complicated graphs 
are involved in case of ${\eta}_c$ production. So, we leave 
these two processes for future investigations.

In Section 2 a short description of the quark-diquark model for exclusive 
processes involving the proton is presented. The details of calculation 
of the hard scattering amplitude for $\Phi$-photoproduction is given in 
Section 3. In Section 4 we give our numerical results and concluding remarks.
\bigskip

\section{ The quark-diquark model}

Let us first notice that,
formally, the amplitudes of the two
processes $\gamma + p \rightarrow V + p$ and $V \rightarrow p + {\bar p}
+ \gamma$ are just related by crossing. Two of us already 
studied the decay process $J/\Psi \rightarrow p+{\bar p}+\gamma$, using 
the quark-diquark model structure of the proton. So, we largely 
refer to our previous paper \cite{16} for notations.

The formalism we are using below is the same as that of (BFL), except that 
the three-body structure of the proton is replaced by a two-body one. To 
lowest order in QCD, the photoproduction of $\Phi$ on proton is thus 
described by the generic graph of Fig. 1. The corresponding amplitude 
is obtained here too from a convolution formula 
\begin{equation}
T=K \int [dx][dx'][dy]\displaystyle{\alpha_{\bf s}^2 \over g^2G^2} 
T_{\mu\nu\alpha\sigma}\varepsilon^{\ast\alpha}_{(\phi)}
\varepsilon^{\sigma}_{(\gamma)}I^{\mu\nu}
\end{equation}

where $[dx]=\delta(1-x_1-x_2)dx_1dx_2$, $[dx']=\delta(1-x'_1-x'_2)
dx'_1dx'_2$, $[dy]=\delta(1-y_1-y_2)dy_1dy_2$. 
Here, collinearity of the constituents with the parent hadron is 
\mbox{assumed :} 
$x_1=x$ (resp. $x'_1=x'$) is the four-momentum fraction of the quark inside 
the ingo0ing (resp. outgoing) proton, and $x_2=1-x$ (resp. $x'_2=1-x'$) that 
of the accompanying diquark ; $y_1=y$ (resp. $y_2=1-y$) is the four-momentum 
fraction of the strange quark (resp. that of the strange antiquark) inside 
the $\Phi$-meson. The tensor $T_{\mu\nu\alpha\sigma}$ is the amplitude 
for the subprocess $g g \gamma \rightarrow \Phi$ with two space-like gluons 
having four momenta $g=xp -x'p'$ and $G=(1-x)p-(1-x')p'$ respectively, $p$ 
being the four-momentum of the ingoing proton and $p'$ that of the outgoing 
proton ; $\varepsilon^{\alpha}_{(\phi)}$ and 
$\varepsilon^{\sigma}_{(\gamma)}$ are polarization vectors for the $\Phi$ 
and for the photon respectively ; $I^{\mu\nu}$ is a tensor amplitude 
describing the two-gluon scattering by a quark-diquark system ; $K$ is the 
overall normalization factor :
\begin{equation}
K = \sqrt{4\pi\alpha}e_{\bf s}\displaystyle{{4{\pi}^2 f_{\phi}}\over
9\sqrt{6}}C
\end{equation}

with the color factor $ C=-2/(3\sqrt{3})$ and the $\Phi$ decay constant 
$f_{\phi}\sim 150$MeV.\\

For the sake of consistency of the model, we have neglected the masses 
of the constituents as well as those of the parent hadrons, whenever 
possible. This led us to use the relativistic form of the $\Phi$ wave 
function. Depending on the helicity $h$ of the $\Phi$, it is given by 
\begin{equation}
\Psi_{\phi} = \displaystyle{f_{\phi}\over\sqrt{24}} 
\displaystyle{1\over\sqrt{3}}
\displaystyle{\sum_{\rm{color}}}s\bar{s}
\left\{ \begin{array}{c}
\not\!{P}~ {\phi}_L (y) ~~\mbox{for}~~ h=0 \\
\not\!{P}~ {\not\!{\varepsilon}}^{(h)}_{(\phi)}~{\phi}_T(y)~~ 
\mbox{for}~~ h=\pm 1 
\end{array} \right.
\end{equation}\medskip

where ${\phi}_L(y)$ and ${\phi}_T(y)$ are normalized $y$-distributions 
for, respectively, a longitudinally and a transversally polarized 
meson.

In that approximation, and due to the particular structure of the 
amplitude of the subprocess (odd number of $\gamma$-matrices), 
it appears then that only longitudinal $\Phi$ are produced.

The tensor amplitude $T_{....}$ in (2) is thus simply obtained 
from the amplitude of $3\gamma \rightarrow e^{+} e^{-}$ 
with massless electrons, by removing the coupling 
constant, using appropriate four-momenta and making the substitution 
\begin{equation}
V_{e^{+}}{\bar U}_{e^{-}} \rightarrow \not\!{P}
\end{equation}

The wave functions of mesons have been derived by (CZ) from QCD sum rules 
technics. We here use their longitudinal-$\Phi$ wave function 
\begin{equation}
{\phi}_L(y)=6 y(1-y)\left\{y(1-y)+0.8\right\}
\end{equation}
that can be found in \cite{10}, p. 259.

Allowing for both scalar $(S)$ and vector $(V)$ diquarks, a quark-diquark
proton state corresponding to a proton helicity ``up" or ``down" takes 
on the general \mbox{form :}
\begin{eqnarray}
&|p^{\uparrow\downarrow}>~ \sim~ - f_{\bf s}~\left[~2\phi_1(x) + 
\phi_3(x)~\right]~S(ud)~{u}^{\uparrow\downarrow} 
\pm f_{\bf v}~\left[~\phi_2(x)
\left\{\sqrt{2}~V_{\pm}(ud)~{u}^{\downarrow\uparrow}
\right.\right.&\nonumber\\ 
&\left.- 2~V_{\pm}(uu)~{d}^{\downarrow\uparrow}\right\} 
\left.+ \phi_3(x)\left\{\sqrt{2}~V_{0}(uu)~
{d}^{\uparrow\downarrow} - 
V_{0}(ud)~{u}^{\uparrow\downarrow}\right\}~\right]&
\end{eqnarray}

where : $V_{h}(q_1q_2)$ is an isovector-(pseudo)vector diquark
state made of two quarks having flavors $q_1$ and $q_2$, $h=0,\pm 1$
being its helicity ; $S(ud)$ is the isoscalar-scalar diquark state. \\

The $\phi_{i}(x)$ are normalized wave functions, 
and $f_{\bf s}$ and $f_{\bf v}$ are normalization constants that
may be chosen unequal to allow for various admixtures of scalar and
vector components. Expressions of diquark-gluon couplings have been 
given in refs \cite{17}. In the space-like channel 
$( g + D \rightarrow D')$ these expressions are, 
using obvious notations and omitting color factors as well as coupling 
constants~:
\begin{equation}
( S'~S)_{\mu} = F_{\bf s}~( D_{\mu}+D'_{\mu})
\end{equation}

for a pair of scalar diquarks and

$$(V'_{h'} V_h)_{\mu} = -F_1~( D_{\mu}+D'_{\mu} )
\varepsilon_{D'}^{h'\,\ast}.\varepsilon_D^h+ 
F_2~\left\{(D.\varepsilon_{D'}^{h'\,\ast})\varepsilon_{D\mu}^h
+(D'.\varepsilon_D^h)\varepsilon_{D'\mu}^{h'\,\ast}
\right\}$$
\begin{equation}
 - F_3~(\varepsilon _{D'}^{h'\,\ast}.D)
(\varepsilon_D^h.D')( D_{\mu}+{D'}_{\mu} )
\end{equation}

for a pair of vector diquarks \footnote{We do not consider here a possible 
mixed coupling involving both scalar and vector diquarks, as it is 
commonly expected to give a small contribution.}.\\

The $F$ above are the diquark form factors depending on $Q^2=-G^2=
-(D-D')^2$. A possible parametrization, aiming at describing 
the natural evolution of the diquark model into the usual 
three-quark picture, has 
been proposed by the authors of 
ref [17]. It has the following form : 
\begin{eqnarray}
\displaystyle { F_{\bf s}(Q^2) = \chi { Q_{0}^2
\over Q^2+Q_{0}^2}}&,&
\displaystyle { F_1(Q^2) = 
\chi \left({ Q_1^2 \over Q^2+Q_1^2}\right)^2} \nonumber\\ 
F_2 (Q^2) = (1+k_{\bf v}) F'_1(Q^2) &,&
\displaystyle{  F_3(Q^2) ={ Q^2F_1(Q^2) \over (Q^2+Q_1^2)^2}} 
\end{eqnarray}
\noindent with
\begin{equation} 
\chi = \left\{ \begin{array}{l}
\alpha_{\bf s}(Q^2)/\alpha_{\bf s}(Q_0^2)~~~
{\mbox{for}}~~~ Q^2 \geq Q_0^2 \nonumber \\
\\
~~~1~~~ {\mbox{for}}~~~ Q^2 \leq Q_0^2 \nonumber \\ 
\end{array} \right.   
\end{equation}\medskip

\noindent $k_{\bf v}$ being the anomalous magnetic moment of the vector 
diquark. The value $k_{\bf v}$=1 is commonly assumed. The above-defined
evolutionary picture also induces one to use, for the sake of
consistency, coupling constants of the running form and to set 
${\alpha}^2_{\bf s} = {\alpha}_{\bf s}(-g^2){\alpha}_{\bf s}(-G^2)$, 
with $-g^2=t x x'$, yet restricting ${\alpha}_{\bf s}$ to some maximum 
value $c_1$. Let us remind that setting the factor $\chi$ in diquark 
form factors provides the correct power of $\alpha_s$'s in 
amplitudes at large transfer.
\bigskip

In \cite{16}, we used such a parametrization to fit the proton magnetic form 
factor $G_M$ in the space-like region. Modelling the momentum 
fraction distributions by a wave function of asymptotic form, i.e. taking 
\begin{equation}
\phi_1(x)=\phi_2(x)=\phi_3(x)=\phi_{\mbox{as}}(x)=20x(1-x)^3
\end{equation}
a quite good fit were obtained with the following values of parameters :  
\begin{eqnarray*}
&\begin{array}{cccc}
f_{\bf s} = 40~ \mbox{MeV}~,& f_{\bf v} = 96 ~\mbox{MeV}~,& Q^2_1= 
2~ {\mbox{GeV}}^2~, & Q^2_0 = 2.3 ~{\mbox{GeV}}^2 
\end{array}& 
\end{eqnarray*}
\begin{eqnarray}
&\mbox{for}~c_1=0.3&
\end{eqnarray}

where, as said above, $c_1$ is the maximum allowed value of the 
running coupling constant ${\alpha}_{\bf s}$. Given that success, 
we used the same parametrization for the present calculation. 

\bigskip

\section{ The hard-scattering sub-amplitudes}

As already mentionned, according to the model here used, 
longitudinal $\Phi$ are 
preferentially produced. There are then, a priori, eight dominant 
helicity amplitudes describing the process, which we denote by 
$T^{\lambda\lambda'\Lambda}$ where $\lambda~\lambda'$ and $\Lambda$ are 
the helicities of, respectively, the ingoing proton, the outgoing proton and 
the real incident photon. Thanks to parity and rotational invariance, 
that number in fact reduces to four and we have :

\begin{eqnarray}
&\begin{array}{cc}
T^{\downarrow \downarrow -}=-T^{\uparrow \uparrow +}& 
T^{\downarrow \downarrow +}=-T^{\uparrow \uparrow -} \\  

T^{\downarrow \uparrow -}=T^{\uparrow \downarrow +}& 
T^{\downarrow \uparrow +}=T^{\uparrow \downarrow -} 
 
\end{array}& 
\end{eqnarray}

To be more specific, let us now concentrate on the calculation of the 
amplitude $T^{\uparrow \uparrow +}$. To compute the amplitudes, 
we have chosen for convenience the polarization states of the 
particles according to a ``t-channel 
helicity-coupling scheme'' \cite{18}. In that scheme, the photon helicities 
represent photon spin projections on the ``vertex plane'' defined by 
the $\Phi$ and photon four-momenta, and the photon polarization vectors  
$\varepsilon^{(\pm)}_{(\gamma)}$ are then perpendicular to that plane. 
Similarly, the helicities of the protons are projections of their 
respective spins on the vertex plane defined by their two four-momenta.

From eq. (2) we thus obtain 

\begin{equation}
T^{\uparrow \uparrow +}=K \displaystyle{\sqrt{2t}\over{t^2}}
\int [dx][dx'][dy]\displaystyle{\alpha_{\bf s}^2 \over{x(1-x)x'(1-x')}} 
{\phi}_L(y) {\cal T}^{\uparrow \uparrow +}
\end{equation}

where

\begin{eqnarray}
& \begin{array}{c}
{\cal T}^{\uparrow \uparrow +}=12 \sqrt{s}\displaystyle{{f^2_v}\over{\mu_v}}
F_2(Q^2)(1-x)(1-x')\cot(\theta/2) \times \\
\displaystyle{1\over{y(1-y)}} \left\{ \displaystyle{{(1-y)A}\over{d'}}+ 
\displaystyle{{yB}\over{d}}-
\displaystyle{{y(1-y)C\sin^2(\theta/2)}\over{d d'}}\right\}
\end{array}&
\end{eqnarray} 

$s$ being the center-of-mass energy squared, $\theta$ the $\Phi$ emission 
angle relative to the incident photon direction in the center-of-mass frame 
($\sin^2(\theta/2)=t/s$), $Q^2 \sim t(1-x)(1-x')$, $\mu_v$ the diquark mass 
usually taken equal to 600 MeV ; $d$ and $d'$ are the 
s-quark propagator factors : 

\begin{eqnarray}
& \begin{array}{c}
d=xx'\sin^2(\theta/2)+y(x\cos^2(\theta/2)-x')-i\epsilon \\
       \\
d'=(1-x)(1-x')\sin^2(\theta/2)+(1-y)((1-x)\cos^2(\theta/2)-(1-x'))
-i\epsilon
\end{array} &
\end{eqnarray}
 
where, following the usual prescription, $\epsilon \rightarrow 0^{+}$. 
Finally, $A$, $B$ and $C$ are given by

\begin{eqnarray}
& \begin{array}{c}
A=\phi_3(x') \phi_2(x)((1-x')\sin^2(\theta/2)+(1-y)\cos^2(\theta/2)) \\
-\phi_2(x')\phi_3(x)((1-x)\sin^2(\theta/2)-(1-y)) \\
       \\
B=\phi_2(x') \phi_3(x)(x'\sin^2(\theta/2)+y\cos^2(\theta/2)) \\
-\phi_3(x')\phi_2(x)(x\sin^2(\theta/2)-y) \\
       \\
C=\phi_2(x') \phi_3(x)\left\{(y(1-x)-x(1-x))\cos^2(\theta/2)+x'(1-y)
-x'(1-x')\right\}\\
+\phi_3(x')\phi_2(x)\left\{(x(1-y)-x(1-x))\cos^2(\theta/2)+y(1-x')-
x'(1-x'))\right\} 
\end{array} &
\end{eqnarray}

From eq. (17), it is clear that the kernel eq. (15) has singularities 
within the domain of integration, since the real parts of $d$ and $d'$ 
have zeroes in $x$ respectively located at 
\begin{eqnarray}
& \begin{array}{c}
z_0= \displaystyle{{x'y}\over{y\cos^2(\theta/2)+x'\sin^2(\theta/2)}} \le 1 \\
\\
{\mbox{and}}    \\
\\
z_1= 1 -\displaystyle{{(1-x')(1-y)}\over{(1-y)\cos^2(\theta/2)
+(1-x')\sin^2(\theta/2)}} \le 1 
\end{array} &
\end{eqnarray}
\bigskip

These singularities correspond to one or the two exchanged s-quarks 
going on-shell in the graph of Fig 1. It is important to notice that when 
$x'=y$ the two zeroes coincide and are both equal to $x'$ (or $y$).   
The one-pole terms ($\sim \displaystyle{1\over d}$ or $\sim 
\displaystyle{1\over d'}$) can be treated readily, using the general 
formula  
\begin{equation}
\displaystyle{1\over{u-i\epsilon}} = {\cal P}\left(\displaystyle{1\over u}
\right)+i\pi \delta(u)
\end{equation}

where $\cal P$ denotes the principal value. The two-pole term 
$\sim \displaystyle{1\over{d d'}}$ corresponds to the graph where the 
photon line is sandwiched between the two gluon lines. Setting $r=\Re(d), 
r'=\Re(d')$ ($\Re$ means real part), one gets  

\begin{eqnarray}
& \begin{array}{c}
\displaystyle{1\over{dd'}} = {\cal P}\left(\displaystyle{1\over r}\right)
{\cal P}\left(\displaystyle{1\over r'}\right)-\pi^2 \delta(r)\delta(r') + \\
 \\
i \pi\left\{ {\cal P}\left(\displaystyle{1\over r}\right) \delta(r')+
{\cal P}\left(\displaystyle{1\over r'}\right) \delta(r)\right\}
\end{array} &
\end{eqnarray}

It appears that the product of two delta-functions, which is 
apparently the most singular term, leads in fact to a null contribution.    
This is to be imputed to the fact that the amplitude of a 
fermion-antifermion-vector-meson vertex is zero when all particles are 
massless.

Let us first consider the imaginary part of the full amplitude. It has the 
general form 
\begin{equation}
C_1(x,x',y) \delta(r) +C_2(x,x',y)\delta(r') + C_3(x,x',y)\left\{ {\cal P}
\left(\displaystyle{1\over 
r}\right) \delta(r')+{\cal P}\left(\displaystyle{1\over r'}\right) 
\delta(r)\right\}
\end{equation}

that can be trivially integrated by hand over the variable $x$, yielding 
an expression of the form :

\begin{eqnarray}
& \begin{array}{c} 
\displaystyle{{C_1(z_0,x',y)}\over \alpha} +\displaystyle{{C_2(z_1,x',y)}
\over \alpha'} + \\
      \\
\displaystyle{1\over \alpha'}C_3(x_1,x',y){\cal P}
\left(\displaystyle{1\over r}\right)_{x=z_1}+\displaystyle{1\over \alpha}
C_3(z_0,x',y){\cal P}\left(\displaystyle{1\over r'}\right)_{x=z_0}
\end{array} &
\end{eqnarray}

where $\alpha=x'\sin^2(\theta/2)+y\cos^2(\theta/2)$ and $\alpha'=
(1-x')\sin^2(\theta/2)+(1-y)\cos^2(\theta/2)$. One must be cautious with the 
two last terms as they are source of difficulties in the subsequent 
(numerical) integrations, as now explained. Since 
\begin{equation}
\displaystyle{1\over \alpha'}{\cal P}\left(\displaystyle{1\over r}
\right)_{x=z_1}=\displaystyle{1\over \alpha}{\cal P}\left(
\displaystyle{1\over r'}\right)_{x=z_0}=\displaystyle{4\over{\sin^2\theta}}
{\cal P}\left(\displaystyle{1\over{(x'-y)^2}}\right)
\end{equation}

a double-pole-like term $\sim 1/(x'-y)^2$
appears. However, that ``singularity'' is tempered by zeroes 
of the factors $C_3$ when $x'=y$ ($C_3 \propto (x'-y)$). These zeroes, 
which are of degree one, are reminiscence 
of the already mentionned fact that the amplitude of a 
fermion-antifermion-vector-meson vertex is zero when all particles are 
massless ; and, precisely, the two exchanged s-quarks that are coupled 
to the real photon in the corresponding ``singular'' Feynman graph get 
both massless when $x=x'=y$. Consequently, the (seemingly) double-pole 
reduces to a simple-pole : 
\begin{equation}
(x'-y)~{\cal P}\left(\displaystyle{1\over{(x'-y)^2}}\right) \rightarrow
{\cal P}\left(\displaystyle{1\over{x'-y}}\right)
\end{equation}

In order to manage this fact in a cautious 
way, we have proceeded as follows. First, we split the products of wave 
functions $\phi$ into symmetrical and antisymmetrical parts :
\begin{eqnarray}
& \begin{array}{c}
S_{23}(x,x')= \displaystyle{1\over 2}\left\{\phi_2(x')\phi_3(x)+\phi_3(x')
\phi_2(x)\right\} \\ 
     \\ 
A_{23}(x,x')= \displaystyle{1\over 2}\left\{\phi_2(x')
\phi_3(x)-\phi_3(x')\phi_2(x)\right\}
\end{array} &
\end{eqnarray}
\bigskip

Of course, this operation is useful only when $\phi_2 \neq \phi_3$ and is 
thus inoperant for the symmetrical parametrization used in this paper. 
We just present it here for further applications. The 
imaginary part of the factor in brackets in formula (16) may then be 
rewritten as 

\begin{eqnarray}
& \begin{array}{c}
\pi \delta(x-z_0)\left\{S_{23}(z_0,x') {\cal S}_0 + A_{23}(z_0,x') {\cal A}_0
\right\} + \\
  \\
\pi \delta(x-z_1)\left\{S_{23}(z_1,x') {\cal S}_1 + A_{23}(z_1,x') {\cal A}_1
\right\}
\end{array} &
\end{eqnarray}

In order to maximally reduce the effect of the pseudo-pole $\propto
1/(x'-y)$, we further made the shift $x' = y + (x'-y)$ 
and the appropriate simplifications in all coefficients $ \cal S$ and 
$\cal A$. We then arrived at the more manageable expressions : 
\begin{eqnarray}
& \begin{array}{c}
{\cal S}_0 = \displaystyle{y\over{{\alpha}^2}}\left\{x'(x'-y)\sin^4(\theta/2)
+ y \alpha (1+\cos^2(\theta/2)) \right.+ \\
(1-y)\left(2x'(2x'-1)+\cos^2(\theta/2)\left[(y-x')(2x'-1)-2y^2+y
\sin^2(\theta/2)\right]\right) +\\
-\displaystyle{{(1-y)}\over{\cos^2(\theta/2)}}\left[(x'+y)(2x'-1)+2y^2
\right] + \\
\left. y^2(1-y)(1-2y)\displaystyle{{(1+\cos^4(\theta/2))}
\over{\cos^2(\theta/2)}}
{\cal P}\left(\displaystyle{1\over{x'-y}}\right) \right\}
\end{array}&
\end{eqnarray}

\begin{eqnarray}
& \begin{array}{c}
{\cal A}_0 = \sin^2(\theta/2)\displaystyle{y\over{{\alpha}^2}}
\left\{x'y +\sin^2(\theta/2) (x'-y)^2+x'-y +y(1-y)\cos^2(\theta/2)\right. +\\ 
\left.-\displaystyle{1\over{\cos^2(\theta/2)}}(1-y)(x'+y)
-y^2(1-y)\displaystyle{{1+\cos^2(\theta/2)}\over{\cos^2(\theta/2)}}
{\cal P}\left(\displaystyle{1\over{x'-y}}\right) \right\}
\end{array}&
\end{eqnarray}
\bigskip

The other factors ${\cal S}_1$ and ${\cal A}_1$ are obtained respectively 
from ${\cal S}_0$ and $-{\cal A}_0$ by the simple replacement 
$x'\rightarrow 1-x'$, $y\rightarrow 1-y$.

Let us now turn to the real part of the amplitude. It appears that different 
terms of the same (large) magnitude compensate each other in the domain of 
integration. To cure 
that new difficulty which causes numerical uncertainties, 
we decided to put all the expression on the same 
denominator $rr'$ and, again, to introduce symmetrical and antisymmetrical 
combinations of wave functions so that the real part of the factor in 
brackets in (16) takes on the form 
\begin{equation} 
\left\{S_{23}(x,x'){\cal S}'+A_{23}(x,x'){\cal A}'\right\}
{\cal P}\left(\displaystyle{1\over r}\right)
{\cal P}\left(\displaystyle{1\over r'}\right)
\end{equation}
\bigskip

Then, we set $u=x'-x$, $v=y-x'$, and rewrited the coefficients ${\cal S}'$ 
and ${\cal A}'$ as polynomials in $u$ and $v$. We thus got
\begin{equation}
{\cal S}'=v^3 H_3 +v^2 H_2 + v H_1 + H_0
\end{equation}
\bigskip

with

\begin{eqnarray}
& \begin{array}{c}
H_3 = -4u\cos^4(\theta/2)
+2\sin^2(\theta/2)(1+\cos^2(\theta/2))(1-2x') \\
      \\
H_2=2u(1-2x')(1+2\cos^4(\theta/2))+\sin^2(\theta/2)(1+
\cos^2(\theta/2))(8x'(1-x')-1) \\
      \\
H_1= u^2\sin^4(\theta/2)(1-2x')+2u\sin^4(\theta/2)x'(1-x') +  \\
2u\sin^2(\theta/2)(1-4x'(1-x')) + 2u(6x'(1-x')-1) \\ 
-2\sin^2(\theta/2)(1+\cos^2(\theta/2))x'(1-x')(1-2x') \\
      \\
H_0=ux'(1-x')\left\{(2u-2x'+1)\sin^4(\theta/2) -2(1-2x')\right\} 

\end{array} &
\end{eqnarray}

and 

\begin{eqnarray}
& \begin{array}{c}
{\cal A}'=\sin^4(\theta/2)\left\{-2v^3  +v^2 (1-2u-2x') + \right. \\
\left.v (2x'(1-x')-u^2) +ux'(1-x') \right\}
\end{array} &
\end{eqnarray}
\bigskip

This trick is supposed to moderate both the above-mentionned cancellations 
and the double pole $\sim 1/(rr')$. At this point, let us notice that 
when $v\rightarrow 0$, one has $1/(rr') \sim 1/(u^2x'(1-x'))$ whereas 
the numerators of the amplitudes behave like $ux'(1-x')$ so that the 
double-pole in $u$ turns into a simple-pole (with no end-point singularities).
Similarly, when $u\rightarrow 0$, then $1/(rr') \sim 1/(v^2 
x'(1-x'))$ while the numerators of the amplitudes are $\sim v x'(1-x')$ so 
that the amplitude is only $\sim 1/v$.

As for end-point singularities due to the denominator 
$x(1-x)x'(1-x')y(1-y)$ coming from gluon and s-quark propagators, they 
cause no problem since the above denominator is cancelled by an 
equivalent factor contained in the product of hadron wave 
functions. We assume that this cancellation of end-point singularities is 
sufficient, namely, that no extra Sudakov form factor that could a priori 
suppress more drastically those singularities is to be implemented. 
The problem of possible Sudakov suppression of end-point singularities is 
beyond the scope of the present work. It has been discussed in different 
contexts by various authors to whom we refer the reader \cite{19}. For our 
calculations, given the uncertainties in the parametrization of the diquark 
model in its present form, we think it senseless to introduce an additional 
complication with a Sudakov factor, the precise form of which for that 
model and for the process here studied is yet unknown anyway. Nevertheless, 
we here apply a kind of suppression of end-point singularities by cutting off 
the dangerous growth of coupling constants $\alpha_s(-g^2)$ and 
$\alpha_s(-G^2)$ in the end-points region by means of the parameter $c_1$ 
(see above).
\medskip

All amplitudes have been treated in the same way. Moreover, to prevent 
additional instabilities in our numerical 
evaluations, we modeled principal values by the approximate form 

\begin{equation}
{\cal P}\left(\displaystyle{1\over z}\right) \sim \displaystyle{z\over 
{\epsilon^2+z^2}}
\end{equation}

with $\epsilon \ll 1$. For instance, in real parts we made the substitution 

\begin{equation}
{\cal P}\left(\displaystyle{1\over{rr'}}\right) \sim \displaystyle{r\over 
{\epsilon_1^2+r^2}}\displaystyle{r'\over {\epsilon_2^2+r'^2}}
\end{equation}

Our numerical results are presented in the next section.
\bigskip

\section{ Results and conclusions}

We spent a lot of time in searching the best method of integration, and 
checking the stability of our computations. Given the rather simple 
dependence on $y$ of the integrant, we tried to integrate by hand 
over $y$ first. This is not a good method because it induces spurious 
singularities 
in the subsequent integration over $x$ and $x'$ through denominators like 
$x \cos^2(\theta/2) - x'$ which does has a zero in the integration domain. 
We also tried a numerical integration in the complex $y$-plane, but this 
also led to intractable instabilities. So, we finally adopted the strategy 
described in Section 3. We then varied the parameters $\epsilon$ down to 
a value as small as 
$10^{-9}$ and even cross-checked our results using two different 
programs of integration, namely a Gaussian quadrature method (RGAUSS) 
and a Monte-Carlo method (VEGAS). It appears that the real parts of 
amplitudes are the most intractable : the less the value of $\epsilon$, the 
greater should be the number of calls of the integrant. However, a value of 
$\epsilon$ between $10^{-4}$ and $10^{-5}$ seems to provide the best 
stability. By chance, for the particular parametrization here used, the 
contributions of real parts appear to be 
much less than those of imaginary parts in the whole kinematical range 
we have investigated, by a factor of a few per cent or less. 
On the other hand, the results obtained for the imaginary parts are very 
stable, being much less sensitive to the value of $\epsilon$.

Thus, the results we are presenting now take account of imaginary 
parts of amplitudes only. 

In Fig. 2, is shown the momentum transfer distribution we obtain in the 
range 2 GeV$^2$ - 10 GeV$^2$ and for two values, 5 GeV (CEBAF) 
and 70 GeV (HERA), of the 
proton-photon invariant mass $W$. One sees that this distribution is 
almost independent of $W$. On the other hand, as expected, the distribution 
exhibits a kind of power-law fall-off, like $t^{-5.5}$ around 3 GeV$^2$ 
and like $t^{-6.5}$ around 9 GeV$^2$. Actually, we have found that the very 
simple form 

\begin{equation}
F(t)=\displaystyle{A\over{t^5 (1+Bt+Ct^2)}}
\end{equation}

with $A=94.5$ nb GeV$^{10}$, $B=-0.113$ GeV$^{-2}$, $C=0.043$ GeV$^{-4}$, 
provides an excellent fit of our results from $t=2$ GeV$^2$ up to values 
of $t$ as large as $15$ GeV$^2$. It may be useful in a generator program 
for a simulation of the process. 
Integrating this form between 
$2$ GeV$^2$ and $10$ GeV$^2$ yields a cross-section of about 1.5 nb. 

Also shown for comparison in the same figure is the fit of 
low-$t$ experimental data 
provided by the ZEUS Collaboration \cite{20}. The observed distribution has 
an exponential fall-off $\sim \exp(-bt)$ with $b \sim 7.3$ GeV$^{-2}$ for 
$<W> = 70$ GeV, in agreement with the expectation of a diffractive 
character of the process in that range. 

The comparison in Fig. 2 is indeed encouraging for the diquark model since 
a direct extrapolation from our results towards lower values of $t$ nicely 
compare in magnitude with the above-mentioned data. Let us remind here 
that according to the well-known asymptotic constituent counting-rule, 
one expects a change of the observed exponential 
fall-off of the $t$-distribution at low $t$ into a power one as $t$ is 
increasing. The diquark model is just supposed to account for that 
transition. One may thus consider the diquark model as a good candidate 
in describing future data at larger $t$, most probably with more 
refined wave functions and diquark form factors, thus establishing, at 
least for that kind of process, a link between diffractive physics 
that holds at low $t$, and the semi-perturbative 
approach of hard exclusive processes one expects to hold at 
very large $t$ (the proton recovering there its three-quark structure).

Fig. 3 shows $s^7 d\sigma/dt$ as a function of $\cos(\theta)$ where 
$s=W^2$. In the pure three-quark picture of the proton, that distribution 
is predicted to be independent of $s$ at large $s$, provided $\alpha_s$, 
the strong coupling constant, is taken as a constant. Obviously, 
that scaling law does not hold here. Essentially, this is because 
we are using 
an evolutionary picture of the diquark structure where 
the $\alpha_s$'s are expressed in running forms. It can be easily 
checked that one recovers the expected scaling law, if one takes 
both the $\alpha_s$'s and the factor $\chi$ in diquark form factors as 
constants. 
However, numerical computations show that this works better for 
$W \geq 10$ GeV. In this respect, $W=5$ GeV is not an asymptotic value.
   
Such deviations of the diquark model predictions  
from the asymptotic scaling law have also been obtained by Kroll 
and co-workers in their recent calculation on photoproduction of $K$ 
and $K^{\star}$ mesons off proton, using the diquark model too with 
a similar parametrization \cite{21}. 

In our numerical calculations, we also tried two other 
parametrizations of the quark-diquark structure of the proton. The 
first one is that of Kroll et al. in \cite{22}. We found that the corresponding 
contribution 
of imaginary parts of amplitudes alone yields a $t$-distribution that is 
larger than that of Fig. 2 by more than one order of magnitude, 
which seems to rule out that parametrization since the corresponding rates 
look too high, especially at low $t$. The second parametrization 
has a quark-diquark wave function derived from the asymetrical three-quark 
proton wave function obtained by King and Sachrajda from QCD sum rules
\cite{23}. 
It is described in \cite{16}. 
This time, the contribution of imaginary parts of amplitudes is much less 
than that obtained from the asymptotic wave function. On the other hand, 
real parts seem now to contribute much more. But, as said before, the latter 
are difficult to evaluate properly because of instabilities in numerical
computations. So, we cannot draw any conclusion at the present time 
about that parametrization. At least, this shows that the 
study of $\Phi$-photoproduction at intermediate $t$ would allow one to 
discriminate between various models \cite{24}. Unfortunately, exclusive 
photoproduction processes are yet largely unexplored at large transfers 
though they possess in that range an undoubted physics potential. 
We hope very much to dispose, in a not too far future, of new data from 
CEBAF or HERA, at least at intermediate transfers.

\begin{figure}[htbp]
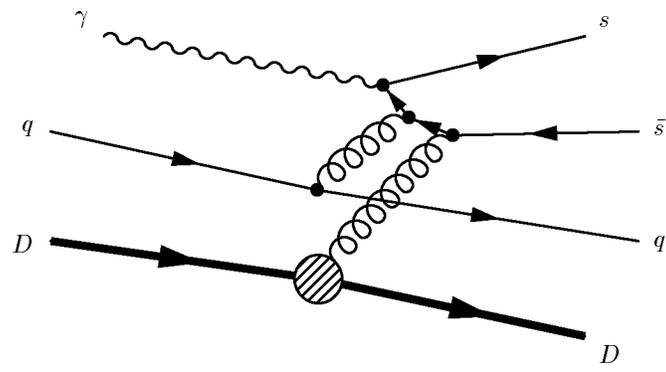

\begin{center}
\leavevmode
\caption{A typical diagram for $\gamma + p \rightarrow \Phi +p$ in the
quark-diquark picture}
\label{fig:1}
\end{center}
\end{figure}
\vskip 0.5 true cm
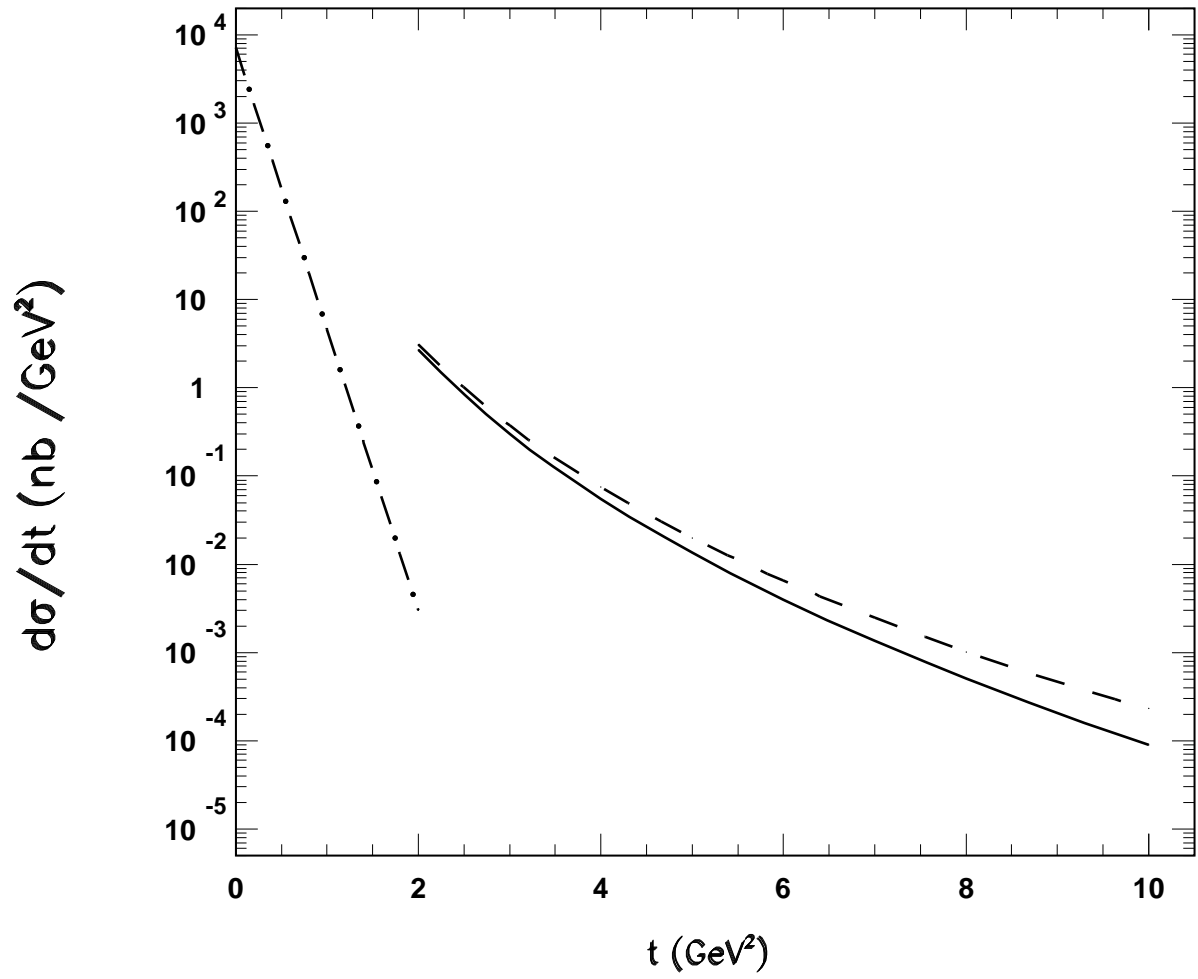
\begin{figure}[htbp]
\begin{center}
\leavevmode
\caption{Diquark-model predictions for $d\sigma/dt$ vs $t$ ; solid line :
$W=5$ GeV ; dashed line : $W=70$ GeV. Dash-dotted line : fit of low-$t$
data at $ <W>=70$ GeV [20]. }
\label{fig:2}
\end{center}
\end{figure}
\vskip 0.5 true cm
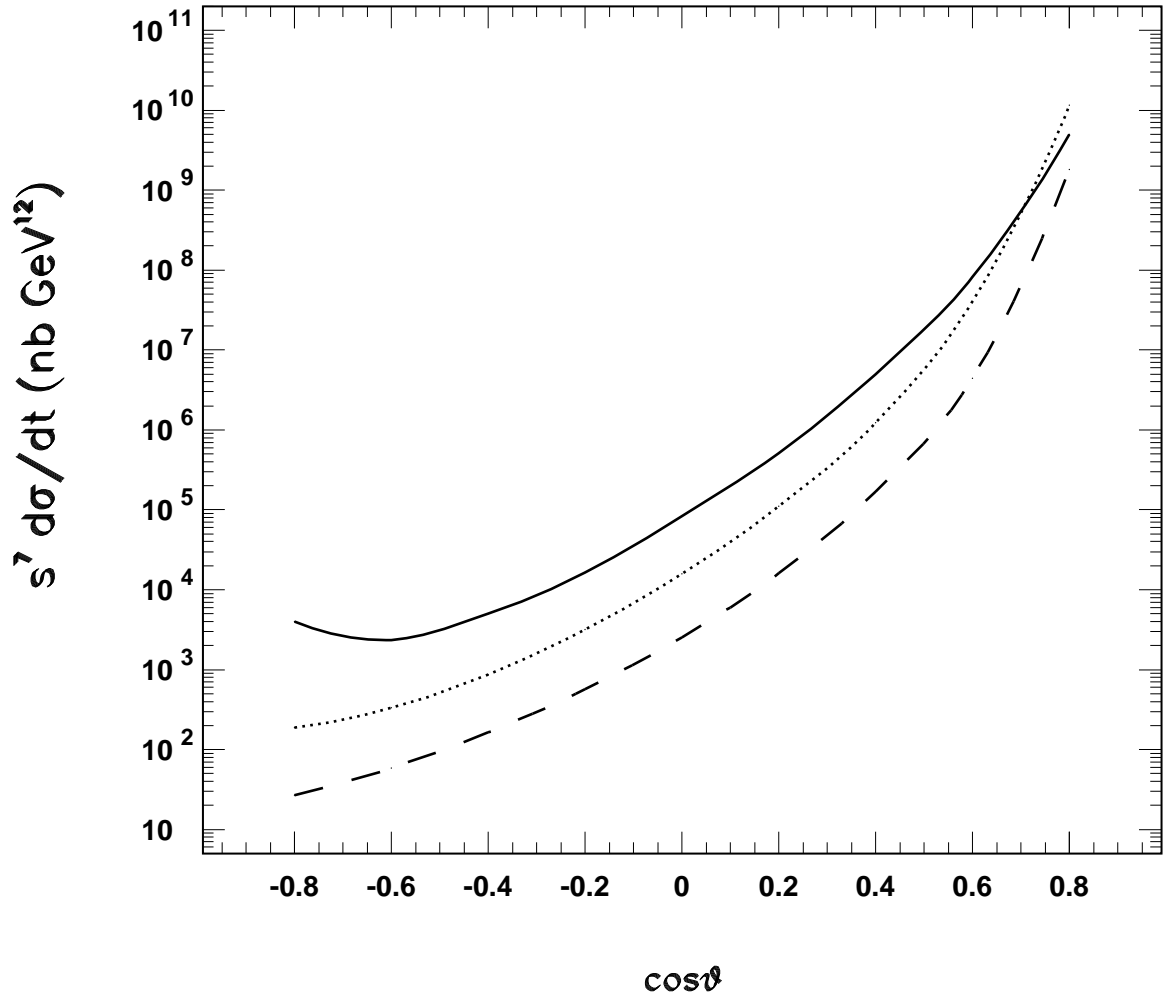
\begin{figure}[htbp]
\begin{center}
\leavevmode
\caption{Diquark-model predictions for $s^7 d\sigma/dt$ vs $ \cos\theta_{cm}$ ;
solid line : $W=5$ GeV ; dotted line : $W=20$ GeV ; dashed line : $W=70$ GeV.}
\label{fig:3}
\end{center}
\end{figure}


\begin{references}
\bibitem{24} One may wonder at this 
point whether our calculation is sensitive or not to the choice of the 
$\Phi$ wave function. As results from our numerical computations, 
replacing the (CZ) wave 
function by the asymptotic one ($6y(1-y)$) 
increases the rates by about 40$\%$.
\end{references}
\end{document}